\documentclass[conference]{IEEEtran}
\IEEEoverridecommandlockouts
\usepackage{cite}
\usepackage{url}
\usepackage{amsmath,amssymb,amsfonts}
\usepackage{algorithmic}
\usepackage{graphicx}
\usepackage{textcomp}
\usepackage{xcolor}
\usepackage{subcaption}

\def\BibTeX{{\rm B\kern-.05em{\sc i\kern-.025em b}\kern-.08em
    T\kern-.1667em\lower.7ex\hbox{E}\kern-.125emX}}

\begin{document}

\title{GPU backed Data Mining on Android Devices}

\author{\IEEEauthorblockN{Robert Fritze}
\IEEEauthorblockA{\textit{Faculty of Computer Science} \\
\textit{University of Vienna} \\ 
Vienna, Austria \\
ORCID 0000-0001-7061-9587}
\and
\IEEEauthorblockN{Claudia Plant}
\IEEEauthorblockA{\textit{Faculty of Computer Science} \\
\textit{University of Vienna}\\
Vienna, Austria \\
ORCID 0000-0001-5274-8123}
}

\maketitle

\begin{abstract}

Choosing an appropriate programming paradigm for high-performance computing on low-power devices can be useful to speed up calculations. %CPUs on low power devices can have a much lower clock rate than those used in computers for which energy efficiency is not an issue. 
Many Android devices have an integrated GPU and  - although not officially supported - the OpenCL framework can be used on Android devices for addressing these GPUs. 
OpenCL supports thread and data parallelism.
Applications that use the GPU must account for the fact that they can be suspended by the user or the Android operating system at any moment. 

We have created a wrapper library that allows to use OpenCL on Android devices. Already written OpenCL programs can be executed with almost no modification.
We have used this library to compare the performance of the DBSCAN and Kmeans algorithms on an integrated GPU of an Arm-v7 tablet with other single and multithreaded implementations on the same device. We have investigated which programming paradigm and language allows the best tradeoff between execution speed and energy consumption.

%We have tried to implemented a programming framework that allows to execute long running jobs on Android devices as background tasks even if the main activity is suspended or the device is rebooted. This is not easy due to restrictions imposed by the operating system. 

Using the GPU for HPC on Android devices can help to carry out computationally intensive machine learning or data mining tasks in remote areas, under harsh environmental conditions and in areas where energy supply is an issue.

%*CRITICAL: Do Not Use Symbols, Special Characters, Footnotes, 
%or Math in Paper Title or Abstract.

\end{abstract}


\section{Introduction}

Low-power CPUs (Central Processing Units) have become
increasingly popular in recent years, especially for mobile or edge computing and if energy efficiency is an issue. These CPUs often have a lower clock rate and, therefore, efficient parallelization techniques may be important to optimize runtime for more complex workloads.% \cite{MontBlanc}.

Android (Open Handset Alliance, Mountain View, USA) is a widely used Linux-based open-source operating system for mobile devices. Android Studio (Google, Mountain View, USA) provides a free and easy to use IDE (integrated development environment) for the creation and analysis of applications. 
Programs written in Java (Sun Microsystems, Santa Clara, USA) or Kotlin (JetBrains, Prague, Czech Republic) are compiled on a host machine and afterwards the executable binary is transferred to the device. %Android supports only a limited set of processor architectures (currently Arm-v7, Arm-v8, x86, x86\_64).
Many Android devices are equipped with an integrated GPU (Graphics Processing Unit) that supports task and data based parallelism. 

The first GPUs have been developed in the 1970s. The purpose of GPUs is to lighten the load of the CPU especially for graphic intense applications (e.g., video games).
Today GPUs can be located directly on the motherboard (integrated GPU) or can be provided through a separate video card (discrete GPU). Integrated GPUs use the main memory whereas discrete GPUs have an own memory. For these GPUs parts of the memory have to be copied to the device before the GPU can transform the data.
Android devices usually have integrated GPUs. 

The OpenCL (Khronos Group, Beaverton, USA, \cite{opencl}) project provides a framework for the compilation and execution of C-style programs on a great variety of platforms like CPUs, GPUs, FPGAs (Field Programmable Gate Array) and many more. 
An advantage of OpenCL is its portability across devices \cite{DU2012391}. Many Android devices are shipped with the necessary OpenCL shared libraries for the integrated GPU on the device \cite{Acosta2018}.

The Android OS (operating system) imposes some restrictions on the program execution that do not apply to usual computing environments. The Android OS is designed to maximize user comfort. Therefore, if the device is not used, the OS activates the standby or doze mode after some minutes to save the battery. Having reached these modes, most activities are suspended.
Moreover, on Android devices the user may decide to move to another activity. 
The activity from which the users moves away is immediately suspended and resources are relinquished. Please see \cite{Androidlifecycle} for an explanation of the life cycle of an Android activity. 
These two restrictions make it impossible to execute long running jobs reliably inside an  activity bound to the UI (user interface). 

\subsection{Contributions}

\begin{itemize}
  \item A framework for using OpenCL compliant GPUs on Android devices. This framework includes  locking mechanism to safely abort calculations performed on the GPU when the user has paused/stopped an activity.
  This framework can be applied not only to data mining tasks but also to vast area of appropriately parallelizable problems (e.g., machine learning) and it complies with the restrictions on the use of external libraries imposed by the Android operating system from Android 7 onwards.
  \item Measurement of the wall clock time needed for the calculation of the DBSCAN and the Kmeans algorithms started with several different parameters. Different implementations (single threaded, multithreaded, GPU) with two different languages (Java and C (Dennis Ritchie \& Bell Labs, Murray Hill, USA)) have been compared.
  \item Analysis of the energy efficiency of the implemented algorithms.
\end{itemize}

\subsection{Previous literature}

\subsubsection{Literature related to the use of GPUs on mobile devices}

All publications related to the use of the GPU on Android devices have used the OpenCL framework.

\textbf{2014} First published attempts to use the GPU on mobile devices were made in 2014. That year Ross et al. \cite{Ross2014} published a paper where they compare the calculation of an N-body simulation on different types of graphic cards. One of them was a GPU on a mobile device.
The same year Wang et al. \cite{Wang2013} implemented an image processing application on Android mobile devices with GPUs. This application allowed the removal of objects from images.

\textbf{2016} Acosta et al. \cite{Acosta2016} have developed a framework based on ParallDroid \cite{ParallDroid} to access the GPU of mobile devices. 
The same year Wang et al. \cite{Wang2016} used the Mandelbrot set to evaluate the performance of an Adreno-GPU (Qualcomm, San Diego, USA) on an Android tablet. 

\textbf{2018 } Acosta et al. \cite{Acosta2018} gave a overview of the availability of OpenCL capable GPUs on mobile devices.
Valery et al. \cite{Valery2018} have implemented PCA (principal component analysis) on Android mobile devices using the GPU. 

\textbf{2019} A year later the same group  has published a framework ``TransferCL'' \cite{Valery2019} based on OpenCL that allows to use deep learning on mobile devices using the GPU.
Afonso et al. \cite{Afonso2019} have implemented a tiling optimization on GPUs of Android devices for tuning the execution of different algorithms.
The same year Fasogbon et al. \cite{Fasogbon2019} have implemented a Depth-Map algorithm on a mobile device using the GPU. 

\textbf{2020} In 2020 Wang et al. \cite{Wang2020} have implemented an neural network training algorithm that optimized for the device capabilities. The training of the neural network was made on the GPU or the CPU.

A few other papers have been published (e.g., \cite{Huynh2017}) in which the GPU of a mobile device is used only for inference of pretrained models. 

\subsubsection{Literature related to data mining on mobile devices}

Many papers have been published on how to analyze data generated by mobile phone (e.g., usage patterns) with data mining method, but only a few publications deal with the implementation of data mining algorithms on mobile devices. We do not focus on publications where mobile devices are only used for collecting data and where the process of data mining or model generation is executed on an external server, even if the generated model is sent back to the mobile device on which inference is carried out, because this is not the scope of this paper.

In 2014 Srinivasan et al. \cite{Srinivasan2014} have published a paper where they present an app called ``MobileMiner'' that allows to perform a data mining tasks on mobile devices. The app was designed for the Tizen operating system.

A paper published 2019 by Yates et al. \cite{Yates2019} evaluates the performance of various data mining algorithms on mobile devices. The authors have imported a Weka \cite{weka} jar archive into an Android project. The authors have used publicly available datasets. The DBSCAN and the Kmeans algorithms were not part of their survey.
Short after the same authors have published an article \cite{Yates2019a} on a data mining app (``DataLearner'') developed by them and designed for the use on Android mobile devices. They followed the same approach they described in the former paper. The authors mention that so far very few attempts had been made to introduce data mining on Android mobile devices.

Earlier (2011) there have already been made efforts to port Weka to Android \cite{rjmarsan}, but this project does not seem to be maintained any more.
Recently Android has added machine learning capabilities to the API (Application programmer interface) \cite{androidmi}. Users can chose to use either pretrained models (e.g., image recognition) or have a custom model trained on an external server. In both cases inference is performed on the device.

As of August 2021 we found many books and tutorials on data mining on Google Play but beside the app ``DataLearner'' created by Yates there was just one other app (``Data Mining Calculator'', published in 2020), that offers data mining algorithms on Android devices.

\section{Methods}

\subsection{Program design}

We use the "WorkManager" job scheduling framework provided by Android for the execution of data mining tasks. The tasks are executed in a deferred manner as background tasks but formally are labelled as foreground service.
This job scheduling service allows to continue tasks even after a reboot of the device.

At the very beginning of the execution of the data mining algorithms, the method acquires a "partial wake lock" which allowes the screen to switch off but kept the CPU at full output.
The lock is released when all jobs have been executed.

We use an app with only one activity. When the activity is started, it tries to connect to the WorkManager and searches for a previously submitted data mining job. If there is none, the activity allowes the user to start a new job. If instead there is a background job, the current progress is read out and displayed. Furthermore, the activity tries to find an OpenCL device. If it is not able to detected automatically a GPU, the user can enter the path to the OpenCL shared library on the device manually.

The execution of the data mining tasks continues even if the user leaves the app, but
the submitted background jobs compete for the system resources with other apps (e.g. CPU cores, GPU).
The parts written in C are accessed from Java using JNI (Java native interface) \cite{JNI}.
The GPU can not be used directly from Java but has to be addressed using the C programming language.

We use a shared wrapper library that serves as glue between the program and the native shared OpenCL library on the device.
A loading method of the wrapper library has to be called before every other call to an OpenCL function. This method loads dynamically the native shared OpenCL library on the device. 
If the library can not be found, an error is returned. The library can be unloaded and reloaded at runtime. The OS decides if and when the memory occupied by the native library is freed or not. 
The wrapper library implements all methods of the OpenCL standard. If an OpenCL method of the wrapper library is called before the shared library has been loaded or after it has been unloaded, an error is returned (an error code that is returned by the native method is used). If the library has been loaded correctly beforehand, the method calls are forwarded from the wrapper library to the native library. The symbols of the native library have to be resolved by hand. This is done only if they are needed and immediately prior to the first invocation of the method.
During the build process, the app is linked against the wrapper library. The wrapper library is packed into the Android apk file. 

\begin{figure}[!t]
\centerline{\includegraphics[
  width=\columnwidth,
  keepaspectratio]
  {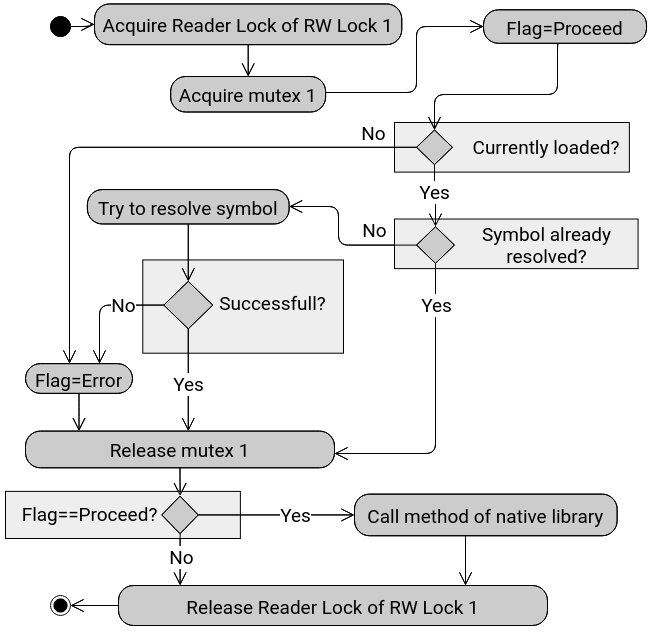}}
  
\caption{Activity diagram of each call to an OpenCL method of the wrapper library.}
\label{fig:1}
\end{figure}

The programmer can use any piece of existing OpenCL code almost without modification. Only the loading mechanism has to be invoked. Besides the loading mechanism, the wrapper library behaves completely transparent with respect to the native OpenCL library. The library is fully thread-safe as long as the underlying native OpenCL library is thread-safe. A reentrant RW (reader/writer) lock is used to get exclusive access to variables that control the loading status of the shared library. Fig. \ref{fig:1} depicts the activity diagram of each OpenCL method of the wrapper library.

The data mining jobs in the background can be stopped by the user (pressing a button in the app). In this case calculations should terminated timely. The app should not wait more than a few seconds to kill the running jobs. Otherwise the display might freeze and a message stating that the app is not responding any more would be displayed. On the other hand, some calculations (especially DBSCAN with many data points) need more than a minute to complete. If the GPU is used, the resources on the GPU have to be freed in an ordered manner. Especially allocated buffers of which the JVM (Java Virtual Machine) is not aware that the GPU still holds a reference, might be subject to premature automatic garbage collection. 

We use a self implemented reentrant writer preferred RW (reader writer) lock. All Kmeans and DBSCAN implementations have to test from time to time a flag and check if they should abort immediately. For the implementations that use the GPU, the flag is tested between OpenCL kernel executions.
The flag is accessed using the reader lock of the RW lock. 
To terminate the calculations prematurely, a special method acquires the writer lock. As the lock prefers the writers, from the moment a writer is waiting, all new readers have to queue up. After the readers, that already have acquired the lock when the writer arrived, have released the lock again, the writer can change the value of the flag.
This mechanism assures that flag is updated by the writer as soon as possible. After the writer has released the writer lock, all waiting readers see the new value.

The source code of the app and the framework is available on Github (Microsoft, San Francisco, USA)  [\url{https://github.com/robertfritze/AGPUDM.git}]. Import the project to AndroidStudio - it should compile and run without any modification. All current Android platforms (Arm-v7, Arm-v8, x86, x86\_64) from API 24 (Android 7) onwards are supported. Please respect the project's MIT licence and the Apache Licence 2.0 of the parts provided by the Khronos group (OpenCL header files).

\subsection{Execution environment}

Please see Table \ref{tab:execenvir} for a description of the execution environment we have used. 
During the execution of the calculations the device was always attached to the power line. Wifi and Bluetooth were switched off. No other activity was used. 

The hardware acceleration has been switched off (developer options of the device and in the app's manifest file). This should preserve the maximum GPU output for the execution of the OpenCL jobs.

We did NOT use a rooted device for this project. Please note, that the device used is an Arm-v8 device, but was used in Arm-v7 mode. This is imposed by the device/the vendor and can not be changed by the user. Arm allows to switch between legacy Arm-v7 and thumb Arm-v7 mode (not used), but does not allow to switch between Arm-v7 and Arm-v8 execution modes (not even on rooted devices). The app has been tested on tablets and mobile phones. It has not been tested for Android TV and Android Wear.

We have used AndroidStudio 3.6.3. 
All C programs were compiled with the Clang 8.0.7 (The LLVM Team) compiler with the {\tt -std=c99} and {\tt -O3} switches. The JVM  version 1.8 was used.

\begin{table}[!t]
\caption{Execution environment}
\begin{center}
\begin{tabular}{ |l|c| } 
\hline
Manufacturer & Samsung$^{\mathrm{a}}$ \\
Model & SM-T510 \\
Android Version & 11 \\
Android API & 30 \\
SoC$^{\mathrm{b}}$ name & Exynos 7885 \\
CPU/GPU manufacturer
& Arm$^{\mathrm{c}}$ \\
\hline
\hline
\multicolumn{2}{|c|}{\textbf{\textit{CPU}}} \\
\hline
Type & Cortex \\
CPU-speed (MHz) & 449-1768 \\
Cores       & 2xCortex A73 + 6xCortex A53 \\
Architecture & Arm-v7$^{\mathrm{d}}$ \\
RAM (MB) available & 2879 \\
\hline
\multicolumn{2}{|c|}{\textbf{\textit{GPU}}} \\
\hline
Type & Mali-G71 MP 2 \\
Max. GPU speed (MHz) & 850 \\
Compute units     & 2 \\
\hline
\multicolumn{2}{l}{$^{\mathrm{a}}$ Samsung, Seoul, South Korea, $^{\mathrm{b}}$SoC=System on Chip, } \\ 
\multicolumn{2}{l}{$^{\mathrm{c}}$Arm Ltd, Cambridge, UK, $^{\mathrm{d}}$Arm-v8 in compatibility mode} \\
\end{tabular}
\label{tab:execenvir}
\end{center}
\end{table}

\subsection{Algorithms}

DBSCAN is a very popular density based data mining algorithm that has been developed by Ester et al. \cite{dbscan}. 
%DBSCAN examines every data point and checks if within a certain radius there are at least a given number of other data points. If enough data points can be found, the points are added to the cluster. DBSCAN needs the radius and a number of neighbours threshold as input parameters. 
We use a non-recursive implementation \cite{Wiki_dbscan} of the DBSCAN algorithm, because it is not possible to use recursion with OpenCL. On the GPU we use two DBSCAN kernels that have almost the same purpose: They examine if a data points is (directly) reachable from a given core point. One kernel is called from the main loop, the other during cluster expansion.

%The Kmeans algorithm \cite{kmeans1,kmeans2} assigns every data point to the cluster center with the smallest euclidean distance. The number of clusters is an input parameter of the algorithm. After the assignment, the cluster centers are updated (center of all points assigned to a cluster). The algorithm iterates over these two steps until the centers do not move any more. With finite precision cycling can occur: A solution of a former step is found again and the algorithm enters an endless loop. The initial cluster centers are selected randomly among all data points. 

We use the Lloyd-Algorithm \cite{Wiki_kmeans} for the implementation of the Kmeans \cite{kmeans1,kmeans2} algorithm.
For OpenCL we use one kernel that calculates in parallel the distance of a point to each cluster center and saves the cluster number with the lowest distance.

The buffers necessary for the execution of the kernels are created with the {\tt CL\_MEM\_USE\_HOST\_PTR} flag set. %This should avoid unnecessary memory copy operations. 
The memory is pinned beforehand to indicate the JVM that this part of the memory must not be relocated. The data buffer has also the {\tt CL\_MEM\_READ\_ONLY} flag set.

For the multithreaded version, seven parallel threads are used (eight cores are available on the device used, one is left for the OS). Each thread is assigned a fixed number of data items. The total number of data items is distributed equally among the the threads. 
For Java, the Thread class is derived, whereas for C the pthread library is used.
Semaphores, locks (JAVA) and AtomicBoolean variables (JAVA) are used to synchronize access to variables and for the program execution.

For both algorithms the memory needed for the execution grows linearly with the total number of data items. For each data item an additional 16 bit variable is used. For Kmeans this variable holds the cluster number the data point is currently assigned to. For DBSCAN the first three bits indicate if the data item has been visited and the density reachability. The other bits are used to store the cluster number (0 equals to noise). The first three bits are deleted before the algorithm finishes. 

We use a fixed set of features (1,2 or 4), a fixed set of generated clusters (2,4,6 or 8) and a fixed set of cluster sizes (128, 256, 512, 1,024 and 2,048). In total 60 different combinations (tuples) of theses three variables are tested. We here present the results for the values selected. However, our approach supports data of arbitrary dimensionality, cluster number and cluster size. The program we use  also allows to generate clusters with unequal cluster sizes.

For each pass a new random dataset is generated. All implementations of both algorithms use the same data. The implementations leave the data unmodified. We  generate normally distributed random data with randomly selected cluster centers and randomly selected variances. Different variances are allowed for each feature (if more than one feature is used). All data items are shuffled randomly before the execution of the data mining algorithms. Before a pass starts, the  sequence the methods are executes is also shuffled randomly.

\begin{figure}[!b]
\centerline{\includegraphics[
  width=\columnwidth,
  keepaspectratio]
  {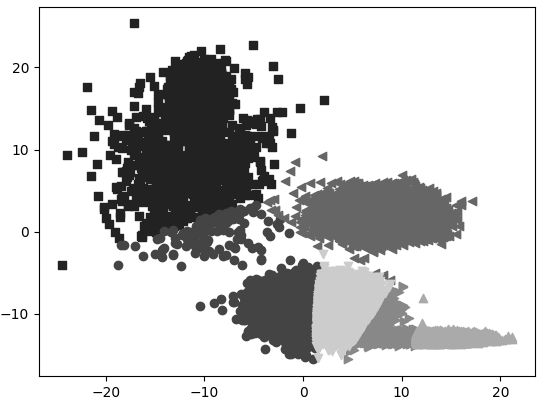}}
\caption{Example Kmeans cluster identification. Same dataset as in Fig. \ref{fig:2} (six clusters generated; 2 features)}
\label{fig:3}
\end{figure}

The Kmeans algorithm has to search as many clusters as have been generated and to terminate if the sum of the absolute displacements of the cluster centers is less than 1e-6 or 100,000 iterations have been completed. The later restriction should avoid endless loops due to cycling which occurs from time to time with single precision. Fig. \ref{fig:3} shows an example of six clusters to be detected by the Kmeans algorithm. 

\begin{figure}[!b]
\centerline{\includegraphics[
  width=\columnwidth,
  keepaspectratio]
  {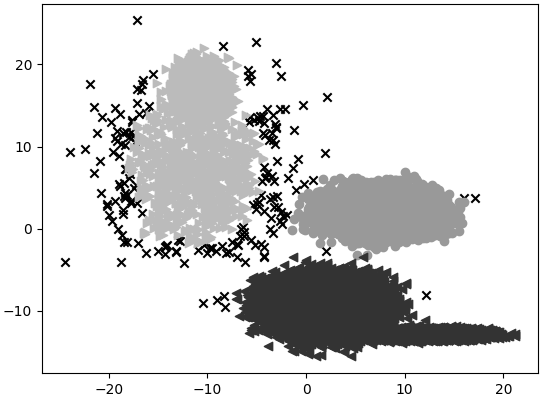}}
\caption{Example DBSCAN cluster identification. Same dataset as in Fig. \ref{fig:3}. (six clusters generated with two features; three clusters found; crosses are noise points)}
\label{fig:2}
\end{figure}

For the DBSCAN algorithm the number of neighbours is set to ten times the number of features and the distance ($\epsilon$) is set to the square root of the number of features.
Fig. \ref{fig:2} shows the result of the DBSCAN algorithm. The same dataset as in Fig. \ref{fig:3} was used.

Both algorithms were implemented in Java (single thread and multithreaded), in C (single thread and multithreaded) and in C with GPU support via OpenCL. All calculations were carried out in single precision.

\subsection{Statistical methods}

We have used R \cite{CRAN} and Python3 \cite{python3} for the statistical evaluation of the results.
We did not perform an outlier detection.
We used the Shapiro-Wilk test \cite{shapiro} to see if data was normally distributed. We have used the Brown-Forsythe-test (R) \cite{brownforsythe} or the Levene-test (Python3) \cite{levene} for the analysis of heteroscedasticity for not normally distributed data. In the case of homoscedasticity we have used the Kruskal-Wallis \cite{kruskalwallis} test or the Wilcoxon \cite{wilcoxon} test to compare distributions. The Bonferroni \cite{bonferroni} correction was applied for the post-hoc analysis of pairwise Wilcoxon-tests. In the case of heteroscedasticity we have used the Median test \cite{moodmedian}. We have used a significance level of $\alpha=0.05$.

The plots were produced with R, Python3 and  VisualParadigm (Visual Paradigm, Hong Kong, China).
For the analysis of the power consumption, data was imported into a SQLite3 \cite{sqlite3} database and processed by Python3.

\section{Results}

In total 4272 tests have been executed on the device. For each of the 60 possible tuples described above, between 65 and 76 tests have been made. For each test the wall clock time for each of the five implementations and both algorithms has been recorded. Additionally we have recorded the time that was used for setting up the threads (in Java and C) and the GPU (buffer allocation, program compilation). In total 16 intervals per test have been stored. 
Furthermore we have tested if all implementations of the DBSCAN algorithm returned the same result. 
For the Kmeans algorithm the initial cluster centers were selected randomly by each implementation and therefore the results were not comparable.

\subsection{Comparison of programming paradigms}
\label{text:compparad}

\begin{figure*}
\begin{subfigure}{.325\textwidth}
  \centering
  \includegraphics[width=0.965\textwidth]{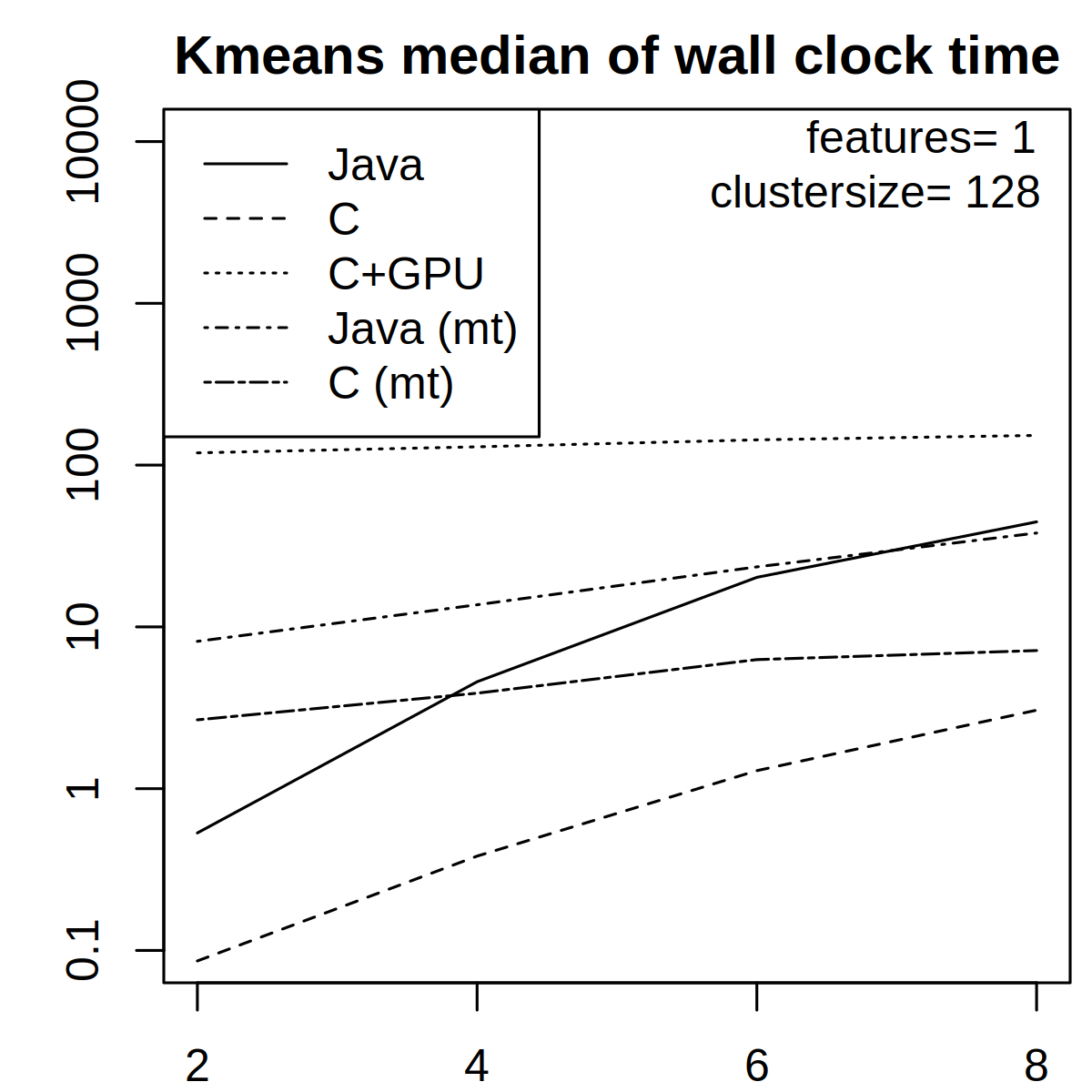}
  \label{fig:sfig11}
\end{subfigure}
\hfill
\begin{subfigure}{.325\textwidth}
  \centering
  \includegraphics[width=0.965\textwidth]{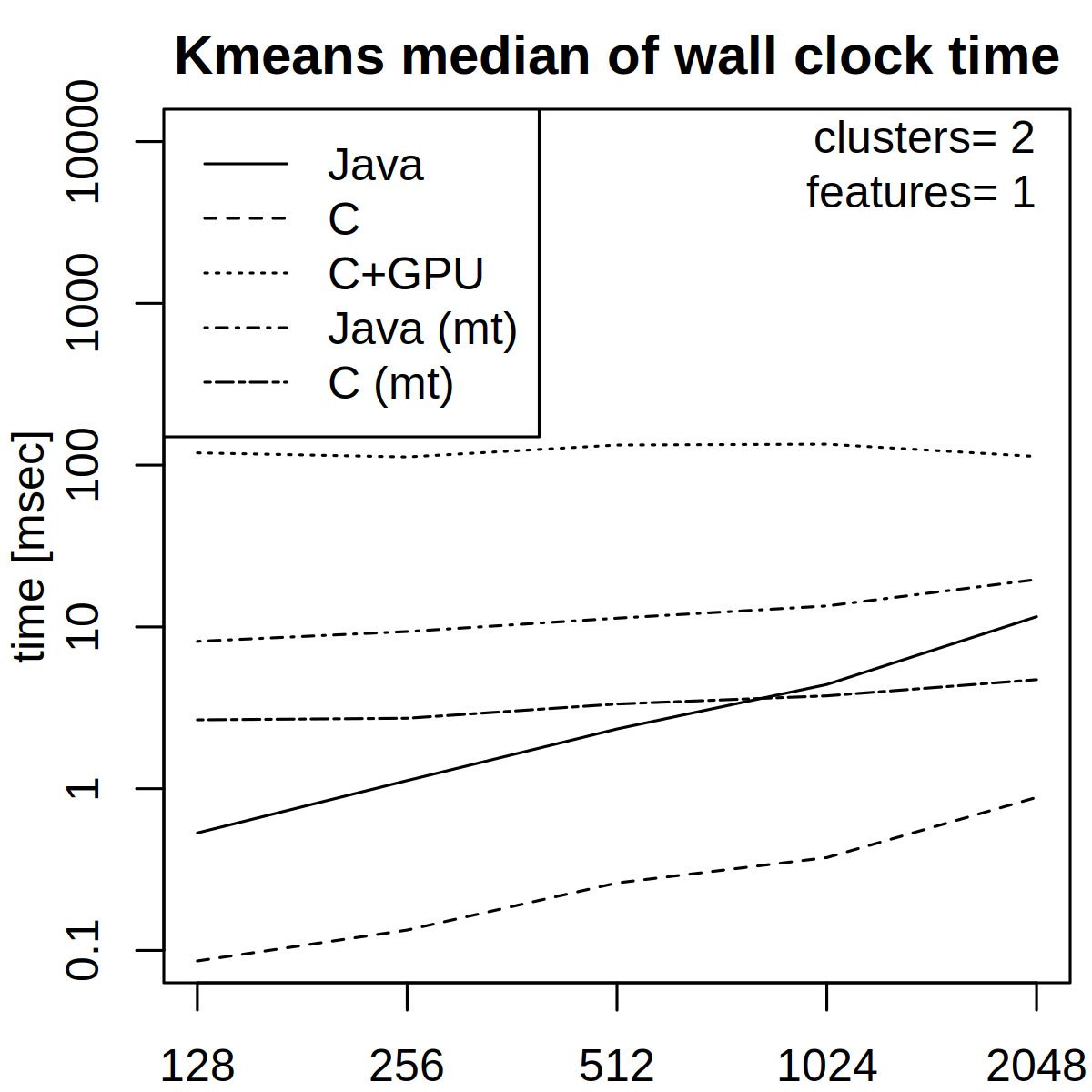}
  \label{fig:sfig12}
\end{subfigure}
\hfill
\begin{subfigure}{.325\textwidth}
  \centering
  \includegraphics[width=0.965\textwidth]{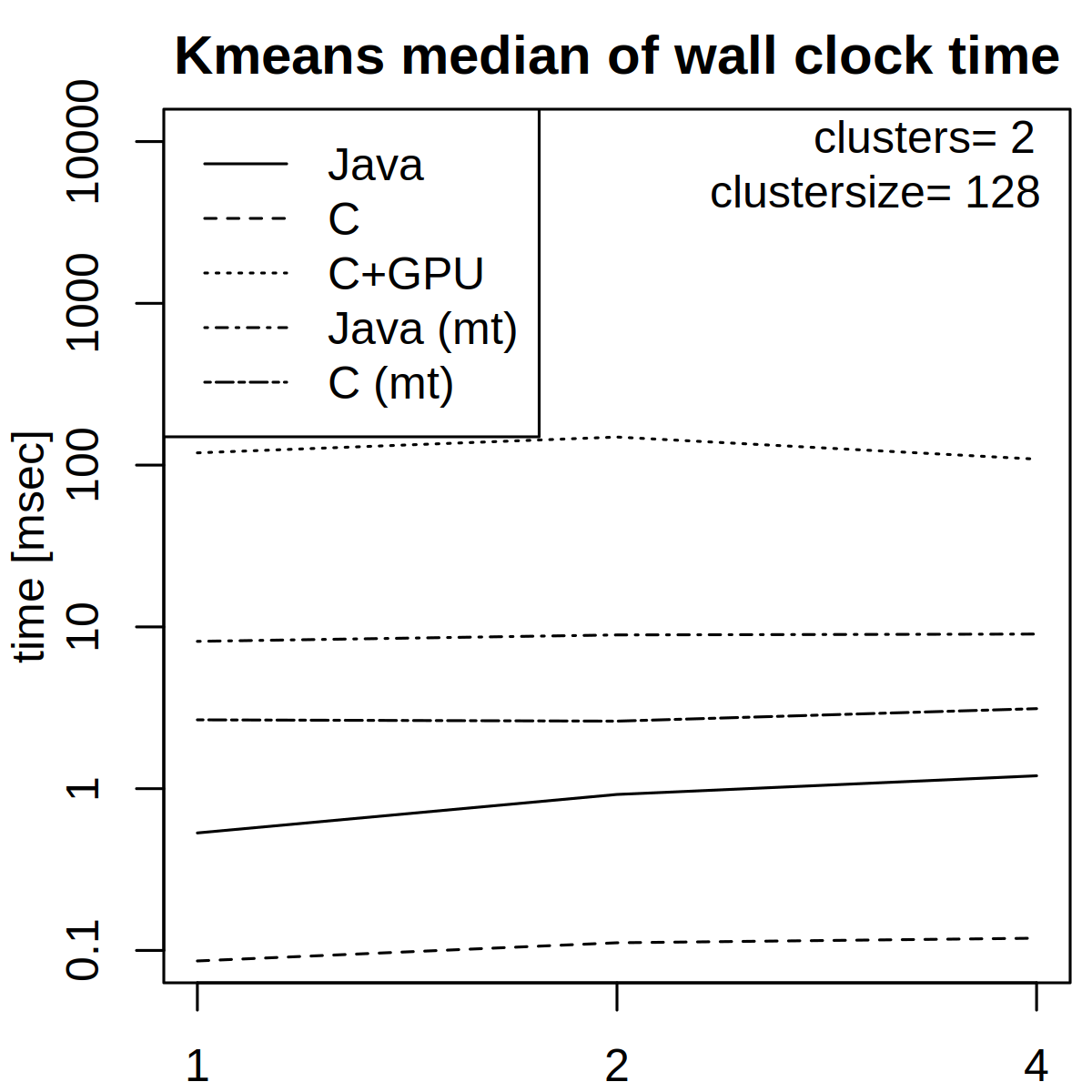}
  \label{fig:sfig13}
\end{subfigure}

\begin{subfigure}{.325\textwidth}
  \centering
  \includegraphics[width=0.965\textwidth]{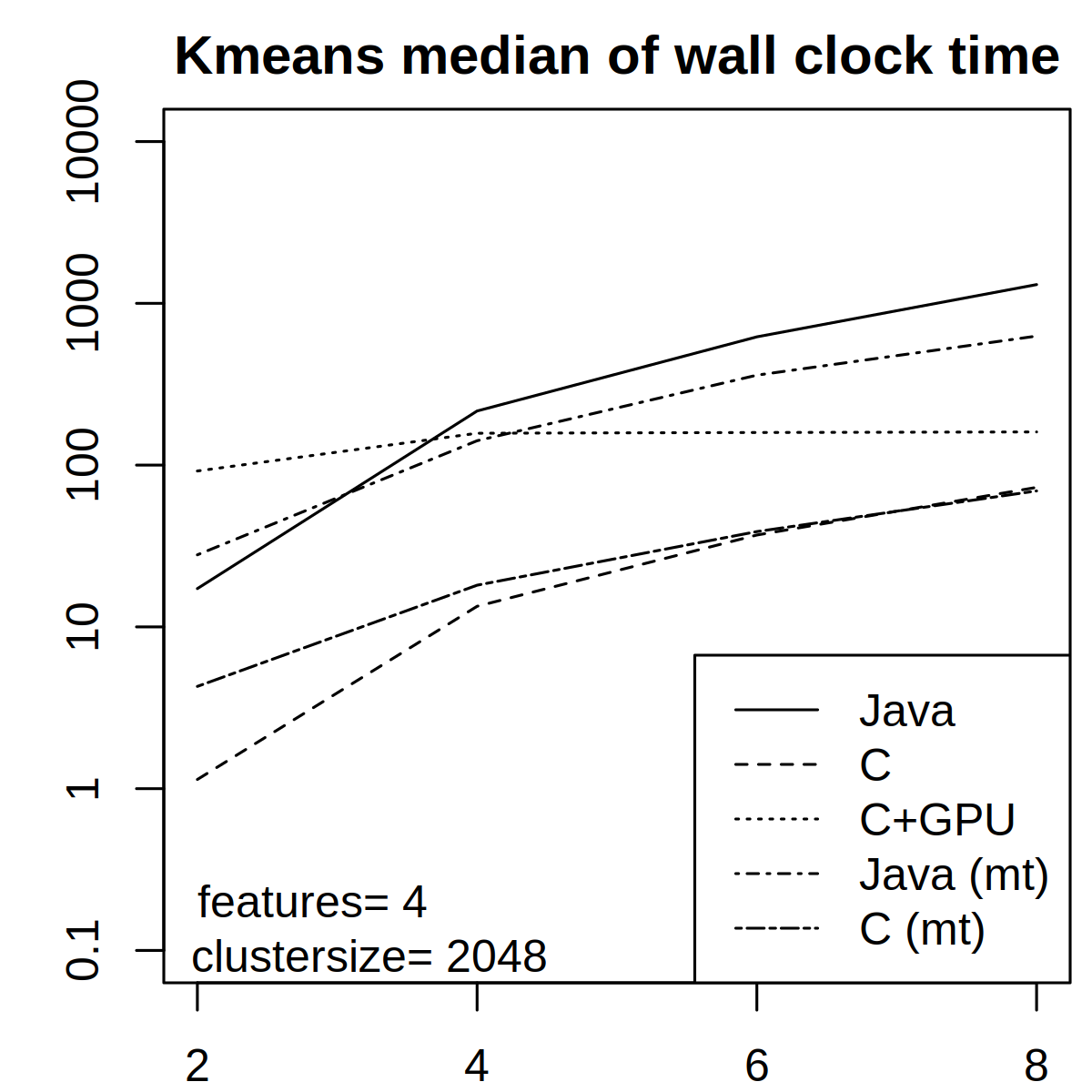}
  \label{fig:sfig21}
\end{subfigure}
\hfill
\begin{subfigure}{.325\textwidth}
  \centering
  \includegraphics[width=0.965\textwidth]{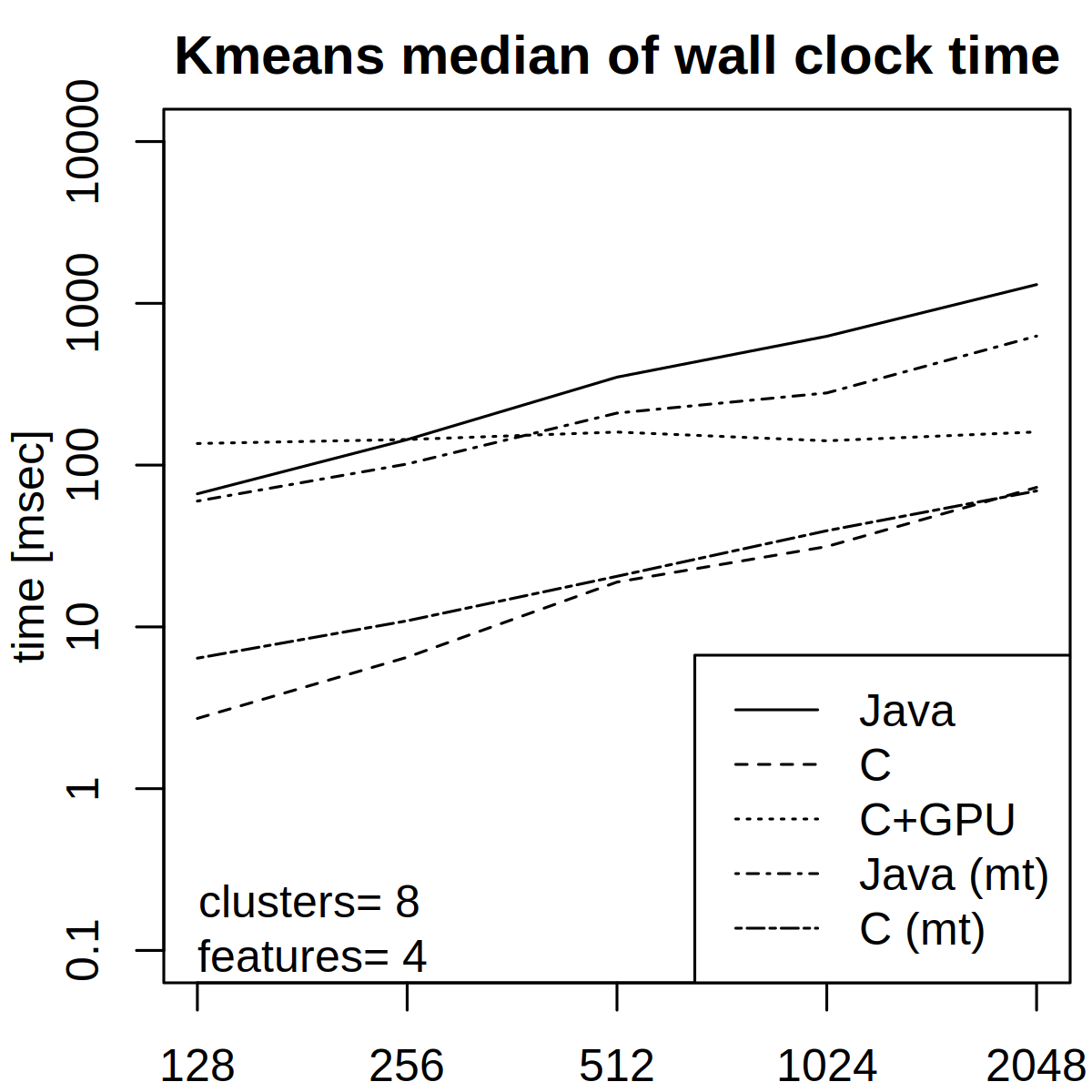}
  \label{fig:sfig22}
\end{subfigure}
\hfill
\begin{subfigure}{.325\textwidth}
  \centering
  \includegraphics[width=0.965\textwidth]{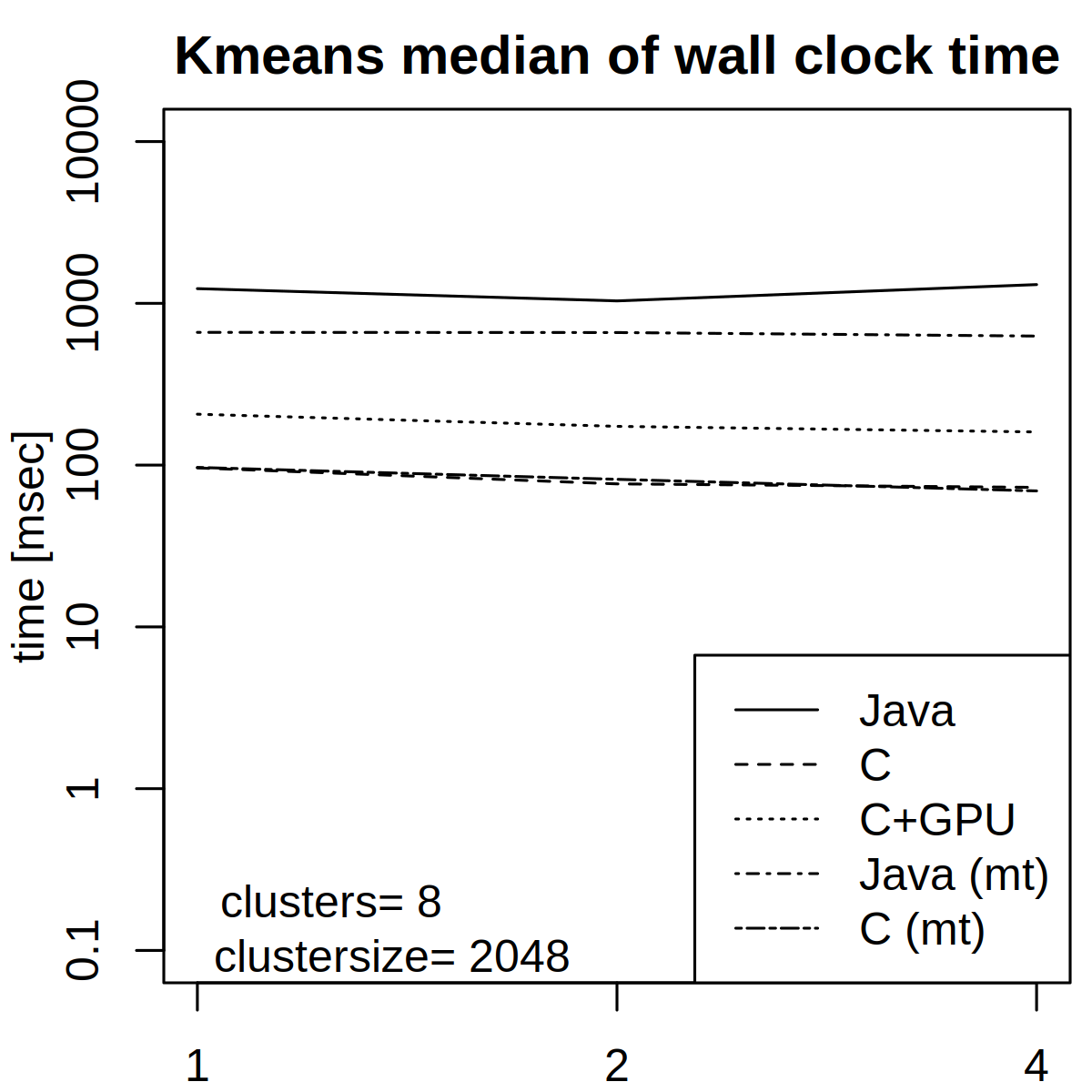}
  \label{fig:sfig23}
\end{subfigure}

\begin{subfigure}{.325\textwidth}
  \centering
  \includegraphics[width=0.965\textwidth]{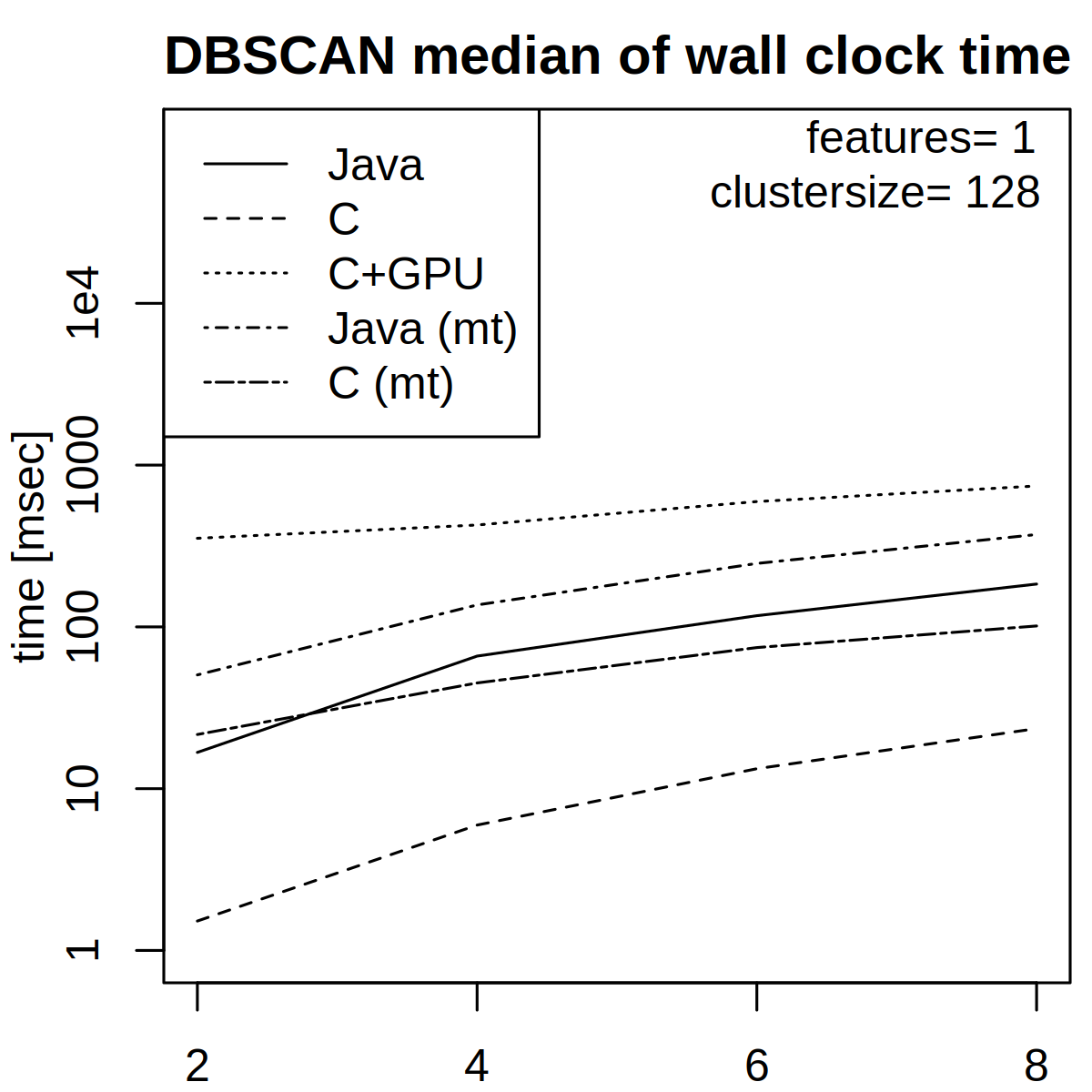}
  \label{fig:sfig31}
\end{subfigure}
\hfill
\begin{subfigure}{.325\textwidth}
  \centering
  \includegraphics[width=0.965\textwidth]{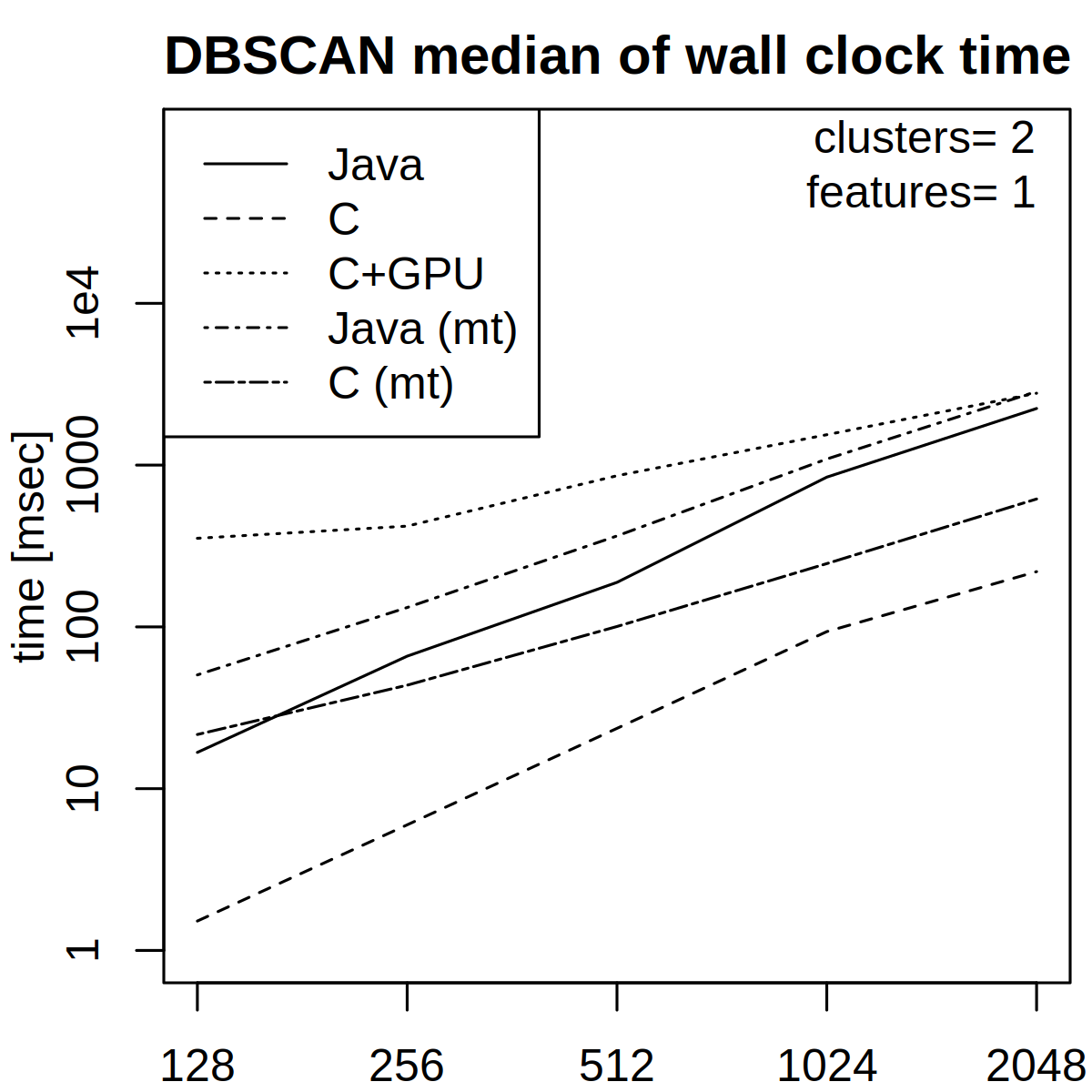}
  \label{fig:sfig32}
\end{subfigure}
\hfill
\begin{subfigure}{.325\textwidth}
  \centering
  \includegraphics[width=0.965\textwidth]{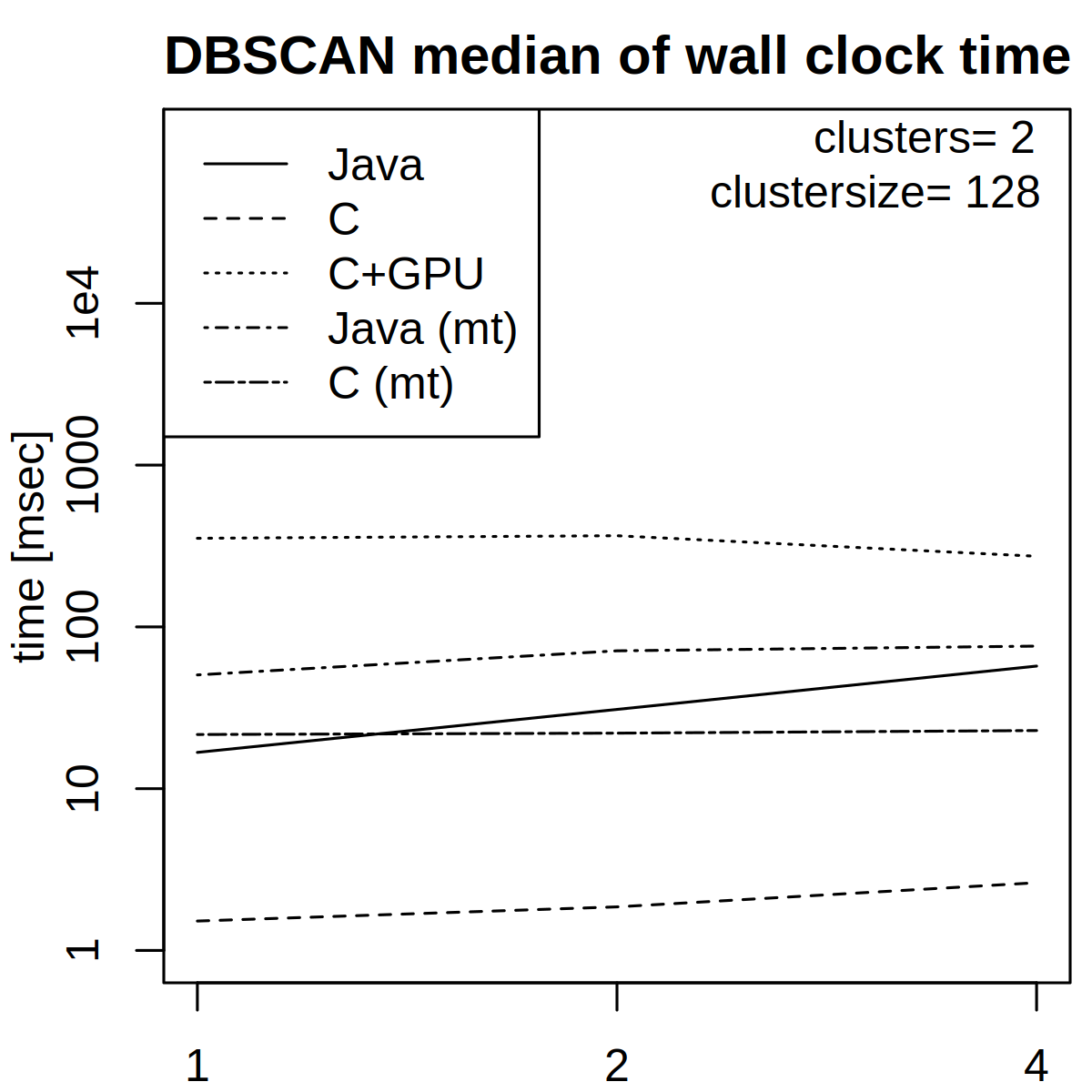}
  \label{fig:sfig33}
\end{subfigure}

\begin{subfigure}{.325\textwidth}
  \centering
  \includegraphics[width=0.965\textwidth]{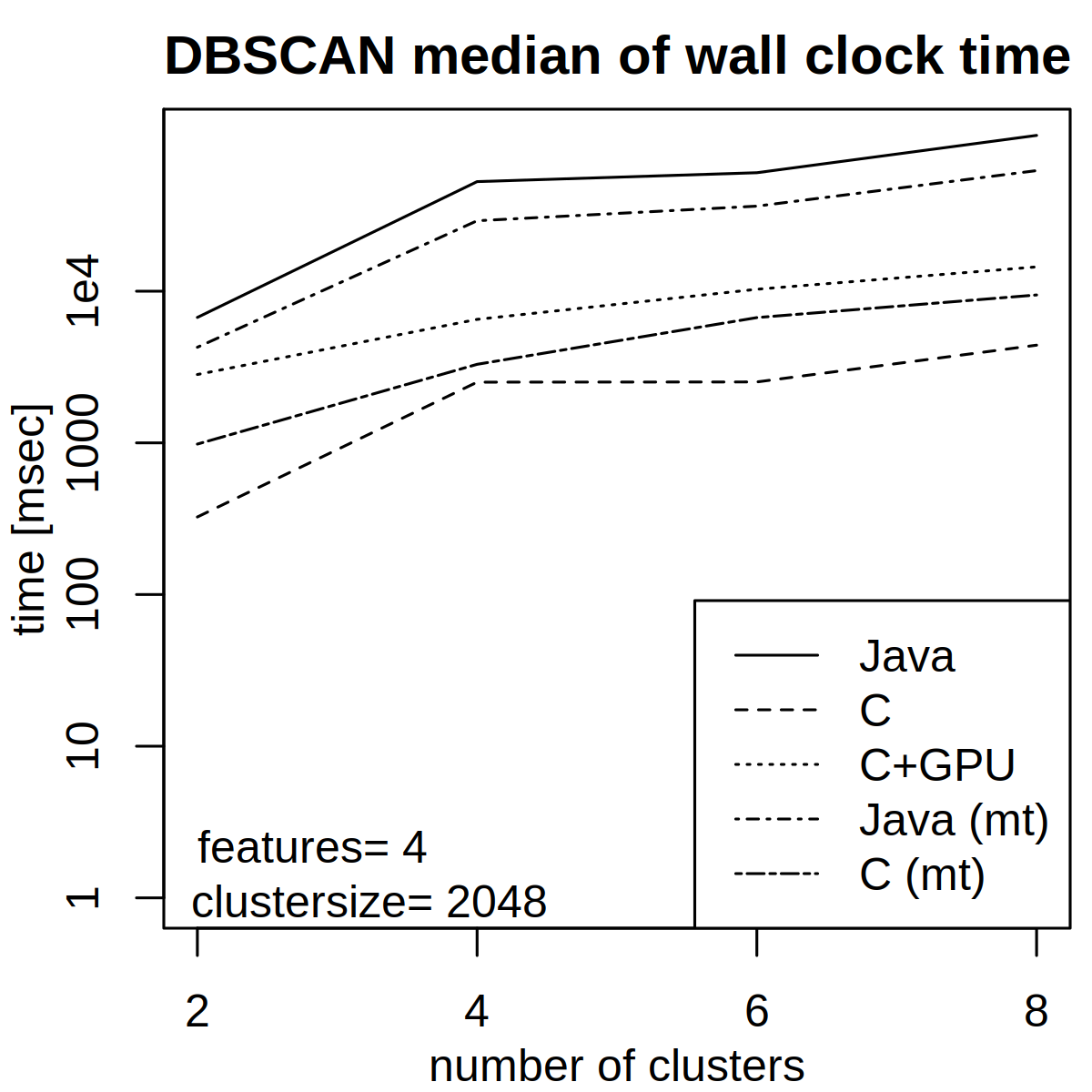}
  \label{fig:sfig41}
\end{subfigure}
\hfill
\begin{subfigure}{.325\textwidth}
  \centering
  \includegraphics[width=0.965\textwidth]{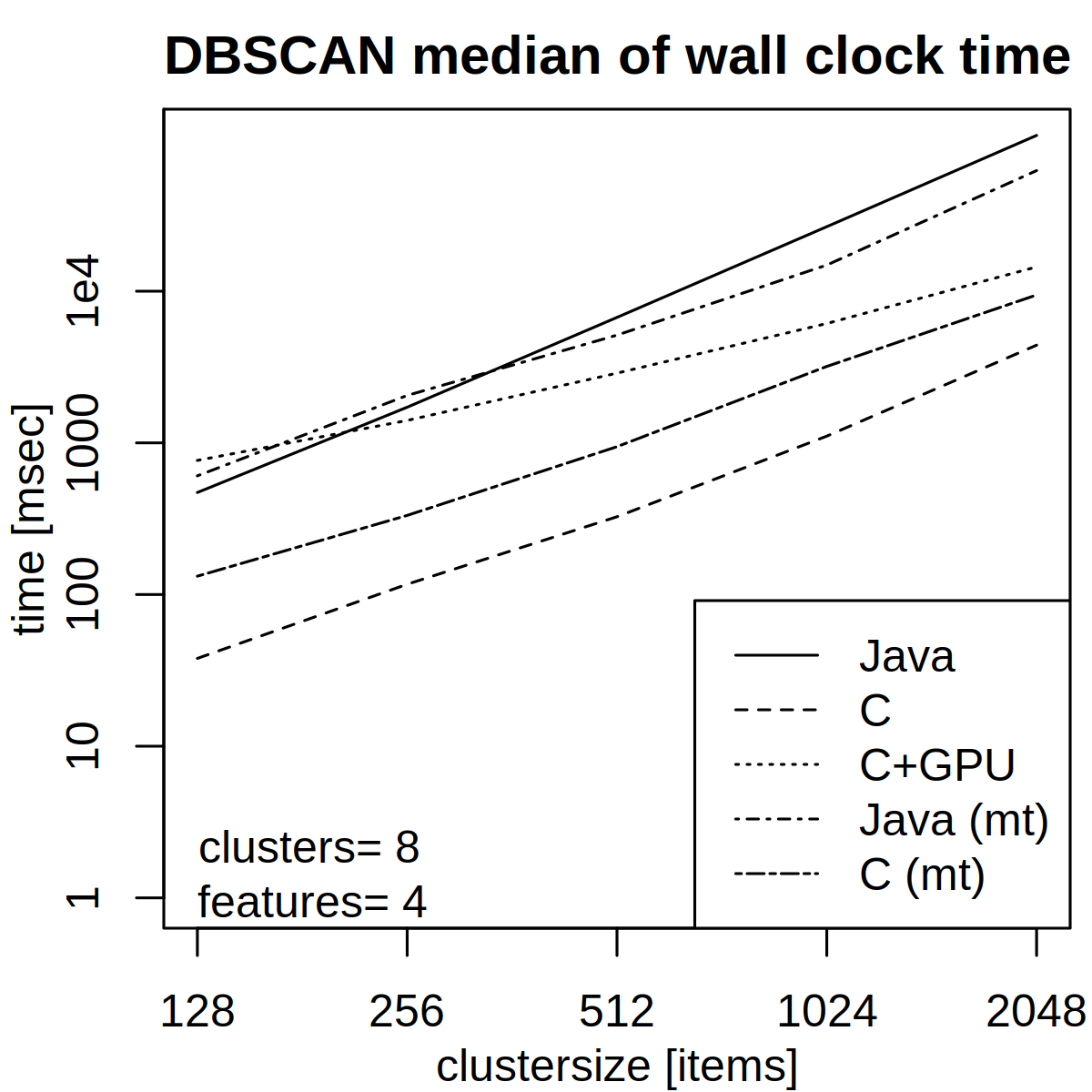}
  \label{fig:sfig42}
\end{subfigure}
\hfill
\begin{subfigure}{.325\textwidth}
  \centering
  \includegraphics[width=0.965\textwidth]{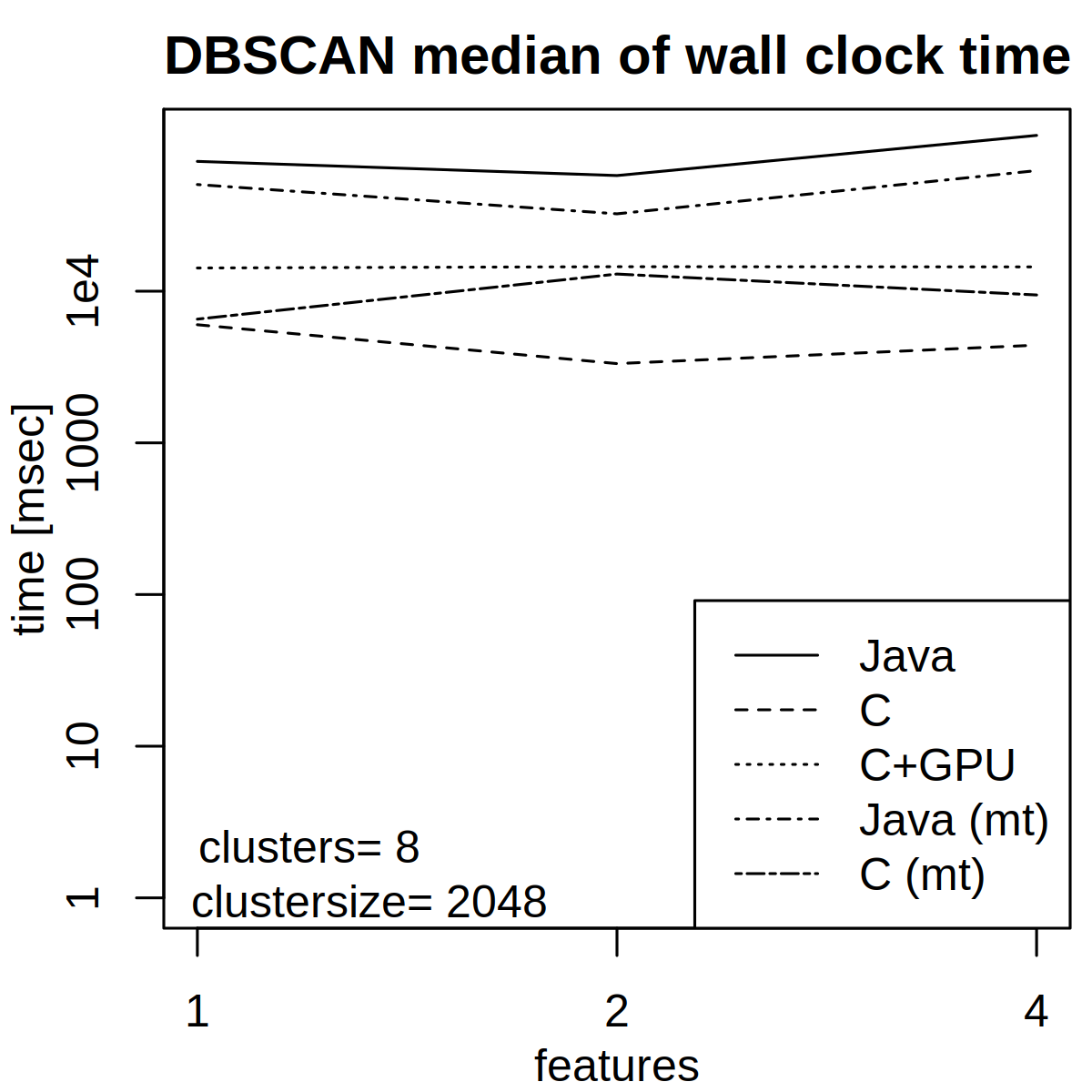}
  \label{fig:sfig43}
\end{subfigure}

\caption{Wall clock time for DBSCAN and Kmeans. Further explanations see \ref{text:compparad}}
\label{fig:figwct}
\end{figure*}

Fig. \ref{fig:figwct} resumes the major findings of the wall clock times (wct) recorded. The upper two rows show the wct for Kmeans, the lower two those for DBSCAN.
The x-axis labels are the same for each column, therefore the x-axis labels are only displayed in the lowermost figures.
For both algorithms the upper row shows the results for the smallest number of data items, the lower row for the largest number of data items. The first column displays the results with respect to the cluster number, the second with respect to the cluster size and the third with respect to the features. Independently of the algorithm we have noticed the following results:

\begin{itemize}
    \item Pure C was always the fastest implementation. 
    \item With longer executions times, Java tends to take longer than all other implementations.
    \item For very short computation times, the GPU takes very long with respect to all other implementations.
    The longer the execution times becomes, the better perfoms the GPU and becomes faster then Java (single and multithreaded).
    \item For very short computation times, multithreaded versions are also not competitive, but they become better the longer the execution takes. Multithreaded C is always faster than multithreaded Java, but multithreaded C is not faster than single threaded C.
    \item The GPU implementation scales better with the number of clusters and the input data size.
    \item Kmeans is much faster then DBSCAN.
\end{itemize}

Please not that in Fig. \ref{fig:figwct} the plots are log-plots (left and right column) or log-log-plots (middle column). 
Both algorithms depend linearly on the number of clusters. Nevertheless, we choose to use a log scale on the y-axsis due to the very large range of values to be displayed. 
For the log-log plots different logarithms have been used on the abscissa (binary logarithm) and the ordinate (common logarithm).
On log-log plots monomials ($y=k\cdot x^\alpha$) appear as straight lines with slope $\alpha$.
If on the abscissa the binary logarithm is used, the slope transforms to $0.301029995664\cdot \alpha$.
Kmeans depends linearly and DBSCAN quadratically on the number of data items.

\subsection{OpenCL overhead}

\begin{figure}[!b]
\centerline{\includegraphics[
  width=0.6\columnwidth,
  keepaspectratio]
  {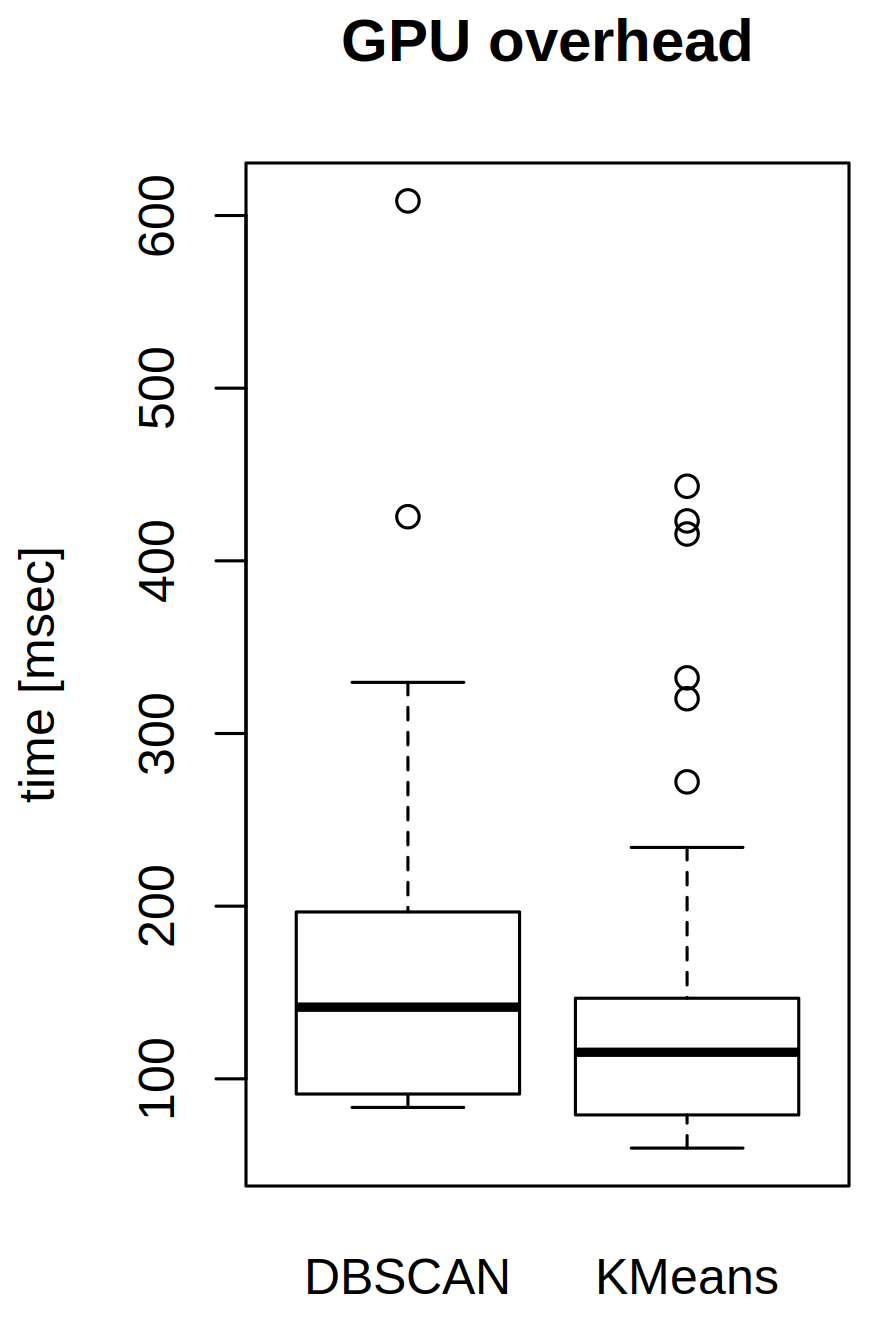}}
\caption{Overhead for GPU setup}
\label{fig:GPUoverhead}
\end{figure}

The use of the GPU implies the overhead of setting up the GPU (buffer allocation, program compilation, etc.). Fig. \ref{fig:GPUoverhead} shows the time needed for setting up the GPU and destroying the buffers afterwards. The time for the execution of the algorithms has been subtracted from the time taken in JAVA before and after the JNI (Java native interface) calls. Therefore, this time includes also the time needed for the JNI.

The difference of the medians is significant ($p<0.05$). The median of the time needed for OpenCL was 141.514ms for DBSCAN and 115.387ms for Kmeans. Sometimes the setup of the GPU takes very long.

\subsection{Thread overhead}

\begin{figure}[!tbh]
\centerline{\includegraphics[
  width=\columnwidth,
  keepaspectratio]
  {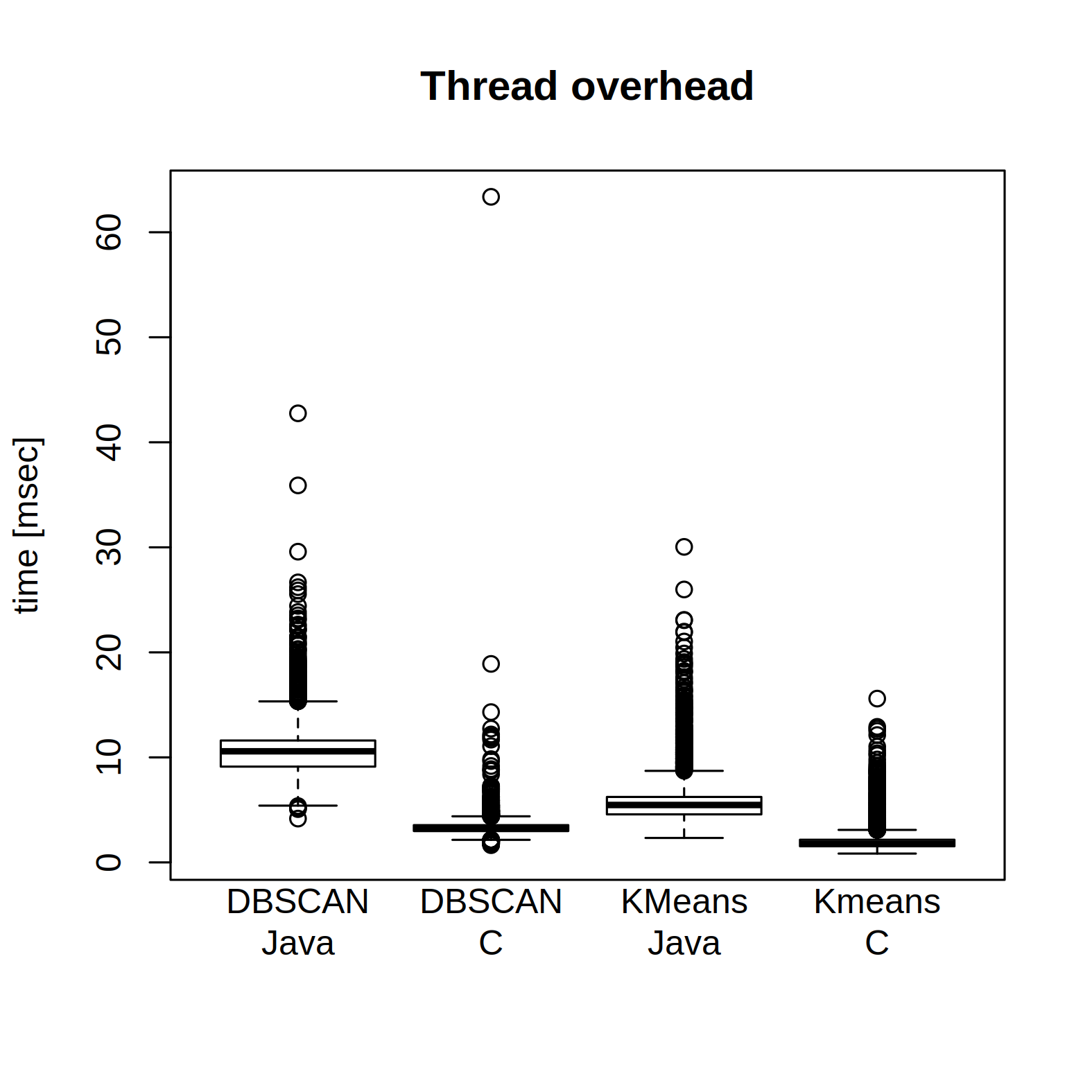}}
\caption{Overhead for the setup and the destruction of threads}
\label{fig:threadoverhead}
\end{figure}

Setting up threads is not for free either, but the time needed is much shorter.
Fig. \ref{fig:threadoverhead} depicts the time needed for the setup the threads. Again this time includes the time needed for the JNI. The intervals are distributed unsymmetrically.
The differences of the medians between Java and C are significant ($p<0.05$) for both groups. 
Table \ref{tab:medianthread} shows the medians of the thread setup times for multithreaded Java and C implementations of DBSCAN and Kmeans. DBSCAN needs approximately twice as much time for the setup as Kmeans.

\begin{table}[htbp]
\caption{Medians for the thread setup times}
\begin{center}
\begin{tabular}{|l|r|r|}
\hline
&\textbf{DBSCAN}&\textbf{Kmeans} \\
\hline
Java & 10.579ms & 5.467ms  \\
C    & 3.198ms & 1.778ms  \\
\hline
\end{tabular}
\label{tab:medianthread}
\end{center}
\end{table}

\subsection{Profiling}

We have used Android's Perfetto profiler that allows to comfortably profile Android devices. The sampeling frequency was set to the lowest possible value (100 or 250ms) depending on the component tested.
The execution time of the Kmeans algorithm was too short to reliably profile the algorithms given the sampling frequency. Therefore, the profiling was made only for DBSCAN. For the profiling tests six clusters, two features and 1,024 data items per cluster were used.  

\begin{figure*}[hbtp]
\centerline{\includegraphics[
  width=\textwidth,
  keepaspectratio]
  {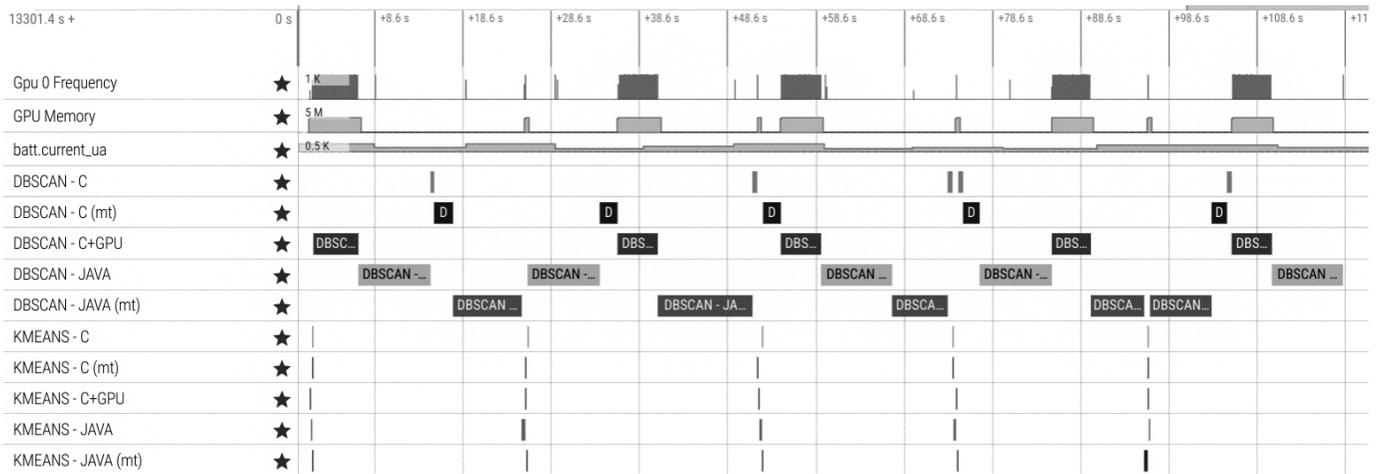}}
\caption{Chart of GPU frequency and power consuption (same trace as in Fig. \ref{fig:cpuusage})}
\label{fig:batgpuload}
\end{figure*}

\begin{figure*}[hbtp]
\centerline{\includegraphics[
  width=\textwidth,
  keepaspectratio]
  {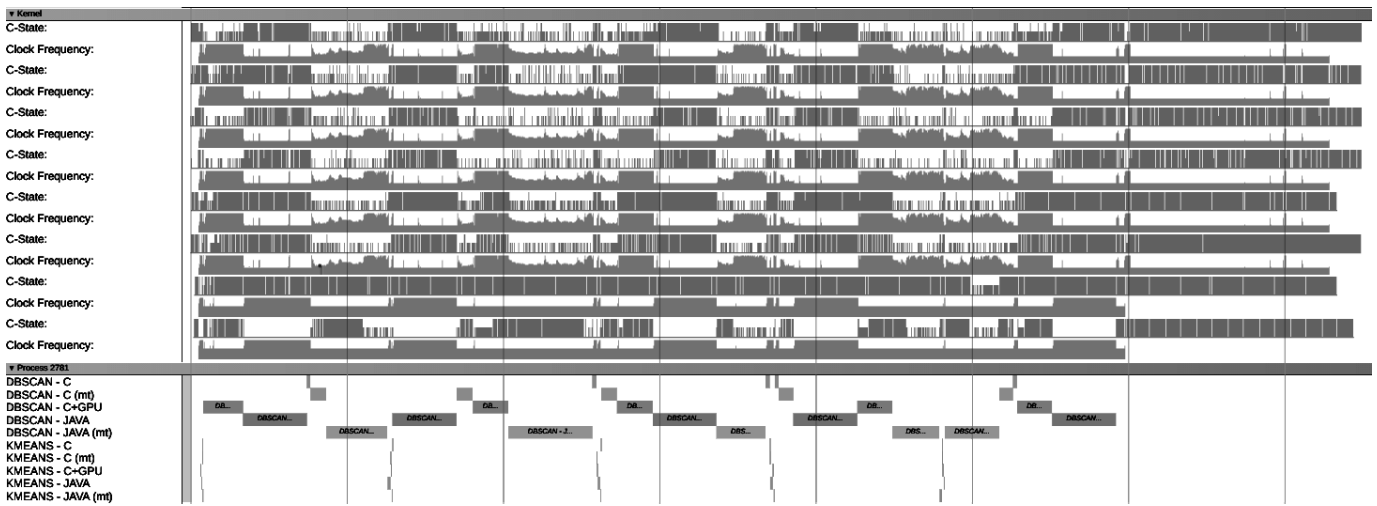}}
\caption{Chart of CPU frequency and C-states (same trace as in Fig. \ref{fig:batgpuload})}
\label{fig:cpuusage}
\end{figure*}

\subsubsection{GPU load}

Fig. \ref{fig:batgpuload} depicts the GPU frequency (``GPU 0 Frequency'') and the GPU memory access (``GPU Memory''). As GPU acceleration has been disabled, the GPU has almost zero background activity and works only if the OpenCL kernels are executed. The GPU kernels require very frequent access to the memory.

\subsubsection{Energy efficiency}

Fig. \ref{fig:batgpuload} shows that there is not much difference in the current while the program is executed. We have investigates this finding more in detail.

We have profiled the counter for the electrical current provided by the profiler. During the profiling the device was attached to the computer (via USB) and therefore the device was charging. The current counter indicates the net current of the device (current out of the battery to the components minus the charging current). The counter can become negative if the charging current is larger than the current consumed by the device. 

\begin{figure}[hbtp]
\centerline{\includegraphics[
  width=\columnwidth,
  keepaspectratio]
  {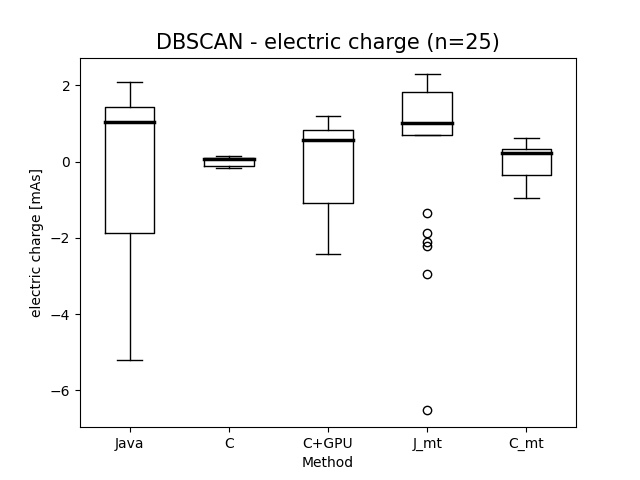}}
\caption{Charge during calculations. J\_mt = multithreaded Java, C\_mt = multithreaded C}
\label{fig:batmedian}
\end{figure}

We have run the test 25 times with the monitor turned on all the time. We have integrated the battery current counter values over the runtime of each algorithm and pass. These integrals gave the electrical charge.
Fig. \ref{fig:batmedian} shows the boxplots of the charge during the calculation for each method. 
The distributions differ significantly ($p<0.05$) and the difference of the medians is also significant ($p<0.05$). 
The median of the charge was highest for Java and lowest for C.
We have analyzed the distributions of the single- and multithreaded Java implementations separately and found no significant difference ($p<0.05$)

We have divided each integral by the runtime of each pass. This gave the mean current during the calculation for each pass. This method respects the individual intervals of the calculations.
There was no statistical significant difference of the distributions and medians ($p=0.85$) regarding the battery current during the calculations.

\subsubsection{CPU frequency and idle states}

Fig. \ref{fig:cpuusage} shows the CPU frequency and idle states. The older UI version had to be used to display the chart. The eight CPUs are shown from the top to the bottom. In the lower part the Gantt chart for the data mining algorithms are shown. The distance between two vertical lines is 20 seconds (a total of 150 seconds have been recorded). A C-state of zero indicates that the core is not idle. Higher values indicate increasing C-states. When a core enters a C-state, its frequency can become unreliable.
Several interesting facts can be observed.

\begin{itemize}
    \item Despite a wake lock is held, the CPUs are allowed to enter C-states. 
    \item After the calculations have finished all cores enter almost immediately a C-state and clock frequency is reduced. 
    \item One core (\#6) is almost all the time idle.
    \item Two cores (\#6 + \#7) exhibit a different pattern than the others. The frequency changes seem to be less frequent and happen on a coarser scale. Furthermore these two cores show always the same frequency, even if one of them is idle.
    \item During multithreaded jobs, seven cores are running but the maximum frequency is rarely attained.
    \item While data mining is performed on the GPU, at least three cores are active.
    \item Single threaded versions of the DBSCAN algorithm are always executed on core \#7 while the others are idle.
\end{itemize}

\section{Discussion}

We have presented a framework including a wrapper library that allows to execute arbitray OpenCL code on the GPU of Android devices almost without modification. We have tested the Kmeans and the DBSCAN algorithms with this framework. We have analyzed the wall clock time of these two algorithms implemented with different programming languages and programming paradigms. %We have shown that the use of the C programming language can significantly speed up both algorithms despite the need of JNI.
As far as we know this is the first publication of a OpenCL framework for Android devices that links to the native OpenCL library on the device only at runtime and not at compile-time.

We wanted to show, that it is possible to reliably perform data mining tasks of moderate size on mobile devices. In contrast to deep learning where complex and large models have to be trained, data mining algorithms implemented in standard programming languages often have a shorter runtime, require less memory and have a smaller energy footprint.
We have selected these two algorithms because we wanted to compare a fast and a slower algorithm and see if it was possible to achieve relevant speedups with alternative programming paradigms. 

Due to the huge effort of setting up the GPU, OpenCL is not well suited for fast algorithms. At least our results suggest that the runtime of the GPU implementation scales better with the workload. Both algorithms need very frequent access to the memory because many distances have to be calculated. %The calculations of a distance matrix beforehand would be possible for DBSCAN but the memory requirement would change from $O(n)$ to $O(n^2)$.
DBSCAN needs more time to set up the GPU than Kmeans because two kernels have to be compiled. 

In earlier published papers the device's OpenCL library was linked directly to the Android binary during the build process (e.g., \cite{Ross2014}, \cite{Wang2016}).
This build process reduces portability significantly, because the APK file has to be compiled for each GPU. 
The authors of \cite{Ross2014} admit that 
``{\it This recipe is complicated and clearly not designed  to  allow  the  easy  execution  of  user  binaries.}''.
On POSIX (Portable Operating System Interface, Austin Group \cite{8277153}) compliant systems, shared libraries can be loaded dynamically and do not have to be present at compile-time. The symbols in the library must be resolved manually at runtime. Android is not fully POSIX compliant but the POSIX standards needed for this project are supported. 
Since Android 7, it is not any more possible to load arbitrary shared libraries on the device at runtime. Vendor provided libraries are an exception to this rule as long as their name appears in specific file on the device.
If they do not, or the list does not exist at all, the shared OpenCL library on the device can not be accessed by the program.

We have used single precision, because integrated GPUs often do not support double precision.
On the other hand, GPUs (even the GPU used for this project) often allow to use half precision. Half precision storage is supported by many CPUs but half precision arithmetics is available only on some Arm-v8 CPUs. 

We used {\tt CL\_MEM\_USE\_HOST\_PTR} flag for the OpenCL data buffers because we wanted to avoid unnecessary memory copy operations. The GPU is an integrated GPU, therefore it will most likely operate directly on the original buffer.
We did not declare the OpenCL data buffer as constant memory because on integrated GPUs there is no difference between global, local and constant memory and OpenCL devices are required to provide constant memory buffers of only 64KB. This limit is too low for data mining purposes.

For future project the following methods could help to exploit the GPU on Android devices better:

\begin{itemize}
    \item Optimize for half precision
    \item Using the same allocated OpenCL memory buffers in multiple different C methods (switching back and forth from C to Java) is perhapes not advisable but one could try to reuse the compiled kernels (either the version compiled for the device or some intermediate representation [SPIR-V]). 
    \item Programs that execute parts of their code on the GPU can use the CPU for other tasks while they are waiting for the results of the GPU.
\end{itemize}

Java multithreaded version performed better than Java except for the case that very few data items had to be processed, whereas multithreaded C implementations were never faster than single threaded ones. This can be explained with the asymmetric CPUs. The single threaded implementations were all executed on the faster and much more advanced (out of order execution) Cortex A73 cores, whereas the multithreaded versions had to use the slower and less powerfull Cortex A53 CPUs. Moreover, these CPUs rarely attained the maximum frequency.
The time needed for the setup of the multithreaded environment takes longer with DBSCAN because this algorithm has to start twice as many threads (one set for the main loop and one set for the cluster expansion) as Kmeans.

%To switch from Java to C JNI was used. This framework has a small overhead should not exceed 200ns (see \cite{criticalnative}).
%In a former unpublished project we have seen that the JNI overhead 
%but according to Android online documentation the time needed should not exceed 150ns. 

We have tried to measure the process and thread time. We have implemented a separate C function that performed a {\tt clock\_gettime} system call with the  {\tt CLOCK\_PROCESS\_CPUTIME\_ID} and {\tt CLOCK\_THREAD\_CPUTIME\_ID} flags set. Besides the fact that is approach adds the JNI overhead to the time measured (Fast- and Critical-JNI calls are available on Android devices only for the operating system) we have got only unreliable values.
We ware not able to use performance counters  in the source code.

We have used the "WorkManager" framework, because jobs can be scheduled in a comfortable way for a deferred execution.
The jobs have to marked as "foreground service" because as long as a foreground service is active, the device will not enter the standby or doze mode (although the screen might switch off).
Furthermore, running as foreground service, tasks are exempted from battery saving policies and are allowed to use more than ten minutes for their completion. The OS imposes that the user has to allow explicitly this behaviour beforehand.
The device was always charging during the calculations because we wanted to preserve the battery life and avoid charging cycles.

%None of the samples drawn was normally distributed. %Outlier present with the Kmeans algorithm can be explained by the fact that cycling occured and the calculations were aborted only after 100,000 iterations.

We have detected no significant difference in the current consumed during the calculations but we have found a significant difference of the charge. We were not able to measure the electric potential of the battery or the power consumed and can therefore not provide direct  measurements of the power. If the voltage of the battery is regarded as constant during the measurement, the electric charge will allow to estimate the power consumed. %The difference of the charge are caused only by the difference of the runtime, not because the power consumed by the CPUs increases when the frequency increases. 

%We expected that a multithreaded version would need much more power to perform the calculations than a single threaded one. 

The sampling frequency for the battery counters was 250ms (lowest possible value). 
%We would have preferred to use Coulomb (As) to measure the energy consumed (integral of the current over time) but due to the presence of a charging current this was not possible. 
The current we have measured includes the current necessary for all components of the device (e.g., monitor). 
We have tried to use the PowerRail configuration to measure the power consumed by the SoC but unfortunately this feature was not present on the device. The device used allows not only to profile it externally (via an USB-cable) but also after detaching it and using the developer options. Unfortunately the device sets the sampling frequency for the battery counters to 1000ms. This frequency  is far to coarse and can not be modified when sampling directly on the device.
The power consumption should be investigated in future surveys with devices that have the PowerRail feature enabled.

The background activity was very low for the GPU (see Fig. \ref{fig:batgpuload}). Hardware acceleration had been switched of beforehand. The background activity of the CPU seemed to be very low as well. Once the calculations have, finished all CPUs enter C-states and the frequencies drop to low values.

Possible future research could focus on the use of SIMD (single data multiple data) instructions. SIMD instructions are able to process several single or double values in parallel (e.g., calculate a distance). 
These days compilers try to extract automatically vector parallelism but sometimes hand coded SIMD assembler is even faster. SIMD instructions can be used with intrinsics (Arm NEON or Intel SSE) or using assembler language together with C. Unfortunately this methods binds the app to the platform the assembler language was written for.
OpenCL provides vector instructions as well. The compiler decides how these vector instructions are translated (true vector instructions or (unrolled) loops).
We have used an Arm Mali Bifrost GPU (Mali G71). These GPUs use quad-style vectorization \cite{quadstylevect} and no explicit vectorization is required. Scalar instructions are executed automatically in parallel.

GPUs are not officially supported by Android. Moreover, not all devices have an OpenCL compliant GPU. If in the future GPUs should become officially supported, for the devices (including the emulators) a workaround must be found. A shared OpenCL library compiled for the CPU (similar to the POCL \cite{pocl} project) could be used. %This would also allow to use vector parallelism on the CPU independently of intrinsics or assembler language.

Executing data mining algorithms on remote devices can help to preserve privacy for sensitive data and supersedes an internet connection.
Regarding the DBSCAN and Kmeans algorithms, the results we have obtained depended on the device used. Currently there are Android devices with more powerful GPUs on the market.

\section{Conclusions}

We have shown that it is possible to use OpenCL on Android devices in a portable and reliable manner. Existing OpenCL programs can be integrated into apps almost without modifications and without rebuilding the app. We have tested two popular data mining algorithms (DBSCAN and Kmeans). %Both tested algorithms have to access the memory very often.
The implementations on the GPU as well as multithreaded implementations suffer from a long setup time with respect to the total execution time. 
%Increasing the workload, the GPU implementation was faster than Java implementations.
%Implementations in C were faster than the same algorithm executed in pure Java. 
On multicore systems with different CPU properties, the runtime of single threaded programs can be better than the runtime of a multithreaded version. We did not find any difference in the mean current consumed during the calculations but the electric charge differed significantly. 
Algorithms that use the GPU (via OpenCL) cloud help to improve execution time except for  algorithms with a very short runtime.

\end{document}